\documentclass[showpacs,amsmath,amssymb,twocolumn,pra,superscriptaddress,notitlepage]{revtex4-1}

\usepackage{qcircuit}
\usepackage{amsmath,bm}
\usepackage[dvips]{graphicx}
\usepackage{amsmath,amssymb,amsthm,mathrsfs,amsfonts,dsfont}
\usepackage{subfigure, epsfig}
\usepackage{braket}
\usepackage{bm}
\usepackage{enumerate}
\usepackage{color}
\usepackage{graphicx}
\usepackage{algorithm}
\usepackage{algorithmic}

\begin{document}


\title{Relationship between costs for quantum error mitigation and non-Markovian measures}


\author{Hideaki Hakoshima}
\affiliation{Research Center for Emerging Computing Technologies,  National  Institute  of  Advanced  Industrial  Science  and  Technology  (AIST),1-1-1  Umezono,  Tsukuba,  Ibaraki  305-8568,  Japan.}

\author{Yuichiro Matsuzaki}\email{matsuzaki.yuichiro@aist.go.jp}
\affiliation{Research Center for Emerging Computing Technologies,  National  Institute  of  Advanced  Industrial  Science  and  Technology  (AIST),1-1-1  Umezono,  Tsukuba,  Ibaraki  305-8568,  Japan.}

\author{Suguru Endo}\email{suguru.endou.uc@hco.ntt.co.jp}
\affiliation{NTT Secure Platform Laboratories, NTT Corporation, Musashino 180-8585, Japan}


\begin{abstract}
Quantum error mitigation (QEM) has been proposed as an alternative method of quantum error correction to compensate errors in quantum systems without qubit overhead. While Markovian gate errors on digital quantum computers have been mainly considered previously, it is indispensable to discuss a relationship between QEM and non-Markovian errors because non-Markovian noise effects inevitably exist in most of the solid-state systems. 
In this work,
we investigate the QEM for non-Markovian noise, and show that there is a clear
relationship between costs for QEM and non-Markovian measures. 
 As examples, we show several non-Markovian noise models to bridge a gap between our theoretical framework and concrete physical systems. This discovery may help in designing better QEM strategies for realistic quantum devices with non-Markovian environments.  
\end{abstract}

\maketitle

\section{Introduction}
It is now widely accepted that quantum computing will enable us to perform classically intractable tasks such as Shor's algorithm for prime  factorization~\cite{nielsen2002quantum}, quantum simulation for quantum many-body systems~\cite{georgescu2014quantum}, and Harrow-Hassidim-Lloyd (HHL) algorithm for solving linear equations~\cite{harrow2009quantum}. However, the effect of decoherence does impose an inevitable impact on the reliability and efficiency of quantum computation, and hence suppressing physical errors is crucial to obtain reliable results~\cite{georgescu2014quantum,hauke2012can,bylander2011noise,reiner2019finding,ma2019dissipatively}.
Although fault-tolerant quantum computing based on quantum error correction can resolve that difficulty, it is not likely to happen for a while because of the large number of physical qubits required per single logical qubit.

Quantum error mitigation (QEM) methods have been proposed to mitigate errors in digital quantum computing, which is compatible with near-term quantum computers with the restricted number of qubits and gate operations as it does not rely on the encoding required in fault-tolerant quantum computing~\cite{preskill2018quantum,temme2017error,endo2018practical,li2017efficient,mcardle2020quantum}. For example, probabilistic error cancellation can perfectly cancel the effect of noise if the complete description of the noise model is given~\cite{temme2017error,endo2018practical}. We apply recovery 
quantum operations $\mathcal{E}_R$
to invert noise processes of gates $\mathcal{N}_G$ such that $\mathcal{E}_R=\mathcal{N}_G^{-1}$. Since an inverse channel of noisy processes is generally an unphysical channel, we need to 
realize
this by applying single-qubit operations with a classical postprocessing of measurement outcomes. Also, the repetition of quantum circuits needs to be $C^2$ times greater to achieve the same accuracy as before QEM, where $C$ is an overhead factor determined by the noise model and operations used in the QEM procedure. We refer to this overhead factor as ``QEM costs'' throughout this paper. We can write QEM costs of the error mitigation of the quantum circuit as $C=\prod_{k=1}^{N_g} c_k$, where $c_k$ is 
the QEM cost
of the $k$th noisy gate and $N_g$ is the number of gates. 
Therefore, in order to suppress costs of QEM, we need to investigate the property of $c_k$ and optimize it.

Recently, \textcite{sun2020practical,endo2021hybrid} proposed a
general quantum error mitigation scheme that can also be applied to continuous quantum systems such as analog quantum simulators. The continuous time evolution of a quantum system is described by 
\begin{align}
\frac{d \rho_N(t)}{dt}=-i[H(t), \rho_N(t)]+ \mathcal{L}\big[\rho _N(t)\big],
\label{eqn:noisyeqn}
\end{align}
where $H(t)$ is the Hamiltonian, $ \rho_N(t)$ denotes a density matrix under
noisy dynamics, and $\mathcal{L}\big[\rho _N(t) \big]$ is the superoperator describing the effect of the environment, which should be mitigated. Even when $\mathcal{L}$ consists of local Lindblad operators, the effects of them easily propagate to the entire system, resulting in highly correlated noise. Note that Eq.~(\ref{eqn:noisyeqn}) can be rewritten as 
$\rho_N(t+ \delta t)=\mathcal{E}_{N} (\rho_N(t) )$, where $\mathcal{E}_{N}$ denotes a superoperator of noisy dynamics
for a small time interval $\delta t$. Thus, similarly to probabilistic cancellation, denoting $\mathcal{E}_I$ as the ideal process, we can apply the recovery channel $\mathcal{E}_R$ such that $\mathcal{E}_R \mathcal{E}_N = \mathcal{E}_I$ using additional single-qubit operations and classical postprocessing of measurement results. In Ref. ~\cite{sun2020practical}, the stochastic QEM method was introduced to implement recovery operations in the limit of $\delta t \rightarrow 0$ via Monte Carlo sampling. Note that  QEM costs corresponding to $\mathcal{E}_R(t)$ can be described as $c(t) \approx 1+ c'(t) \delta t$. Therefore, a QEM cost from $t=0$ to $t=T$ can be described as $C(T)=\exp [\int_0^T dt c'(t) ]$. 


So far, QEM for Markovian noise has been mainly considered and non-Markovian noise has not been investigated well.
The development of the QEM for non-Markovian noise is practically important, because
non-Markovian noise is relevant in most of the solid-state systems such as superconducting qubits, nitrogen vacancy centers in diamond, and spin qubits in quantum dots~\cite{yoshihara2006decoherence,kakuyanagi2007dephasing,bar2013solid,de2010universal,kawakami2014electrical,watson2018programmable}. 

In addition to the practical motivation for non-Markovian noise, 
the concept of quantum non-Markovianity has been extensively studied from a fundamental interest in 
the characterization and the quantification of the backflow of the information from an environment~\cite{rivas2014quantum,breuer2016colloquium}.
There are some definitions of non-Markovianity such as the semigroup definition~\cite{breuer2002theory}, the divisibility definition~\cite{wolf2008dividing,rivas2010entanglement}, and that proposed by Breuer, Laine and Piilo~\cite{breuer2009measure} (Ref.~\cite{rivas2014quantum} discusses a hierarchical relation between these definitions). Throughout this paper, we adopt divisible maps as Markovian processes and we will show that precise definition in a later section.

Moreover, there are many applications to utilize non-Markovianity 
in a positive way
for quantum information processing, including quantum Zeno effects~\cite{misra1977zeno,itano1990quantum}, dynamical decoupling~\cite{viola1998dynamical}, Loschmidt echo and criticality~\cite{haikka2012non}, continuous-variable quantum key distribution~\cite{vasile2011continuous}, time-invariant discord~\cite{haikka2013non}, quantum chaos~\cite{vznidarivc2011non}, quantum resource theory \cite{wakakuwa2017operational}, and quantum metrology~\cite{matsuzaki2011magnetic,chin2012quantum}.
These motivate researchers to investigate the properties of non-Markovianity.

In this paper, we investigate QEM costs for the case of non-Markovian noise. The stochastic QEM can be naturally applied to time-dependent non-Markovian noise to fully compensate for physical errors.
 We show that QEM costs reduce in the non-Markovian region.
We also find a clear relationship between costs of QEM and 
previously reported non-Markovian measures, decay rate measure~\cite{hall2014canonical}, and Rivas-Huelga-Plenio (RHP) measure~\cite{rivas2010entanglement,hall2014canonical}.
We calculate
QEM costs for two experimental setups showing non-Markovianity as examples:
One is a controllable open quantum system, which consists of a long-lived qubit coupled with a short-lived qubit. This system has been 
realized
in NMR experiments~\cite{matsuzaki2019realization,Kukita_2020}.
The other example is a qubit dispersively coupled with a dissipative resonator~\cite{govenius2015parity,blais2004cavity,bertet2005dephasing}.
This discovery may illuminate how to construct efficient QEM procedures.

The rest of this paper is organized as follows.
In Sec.~II, we review the stochastic QEM proposed in Ref.~\cite{sun2020practical}.
In Sec.~III, we review the definition and the measure of non-Markovianity.
In Sec.~IV, we discuss the relation between QEM costs and the measure of non-Markovianity, and study QEM costs for specific models.
Finally, we summarize and discuss our results in Sec.~V.

\section{Preliminaries: Stochastic quantum error mitigation}
In this section, we review the stochastic QEM~\cite{sun2020practical}. Suppose that
the dynamics of the system of interest can be described by Eq.~(\ref{eqn:noisyeqn}). Here, we assume that the local noise and the coupling to the environment is sufficiently weak and the continuous dynamics of the system can be described by the time-dependent Lindblad master equation.
Now we express the evolution of the state from $t$ to $t+\delta t$ as 
$\rho(t+\delta t)=\mathcal{E}_{N}(t) (\rho(t))$ and $\rho(t+\delta t)=\mathcal{E}_{I}(t) (\rho(t))$, 
corresponding to the noisy process and the ideal process, respectively. 
The ideal process represents the unitary dynamics without any noisy operators in Eq.~(\ref{eqn:noisyeqn}).
We hope to emulate the ideal evolution $\mathcal{E}_{I}$ by mitigating errors of the process, $\mathcal{E}_{N}$. When the evolution is affected by local noise operators, i.e., $\mathcal{L}$ can be decomposed as a linear combination of local noise operators, by using a recovery operation $\mathcal{E}_Q (t)$, 
we can efficiently find a  decomposition: 
\begin{align}
\mathcal{E}_I(t)&= \mathcal{E}_Q(t) \mathcal{E}_N(t), \\
 \mathcal{E}_Q (t)&=\sum_i \mu_i \mathcal{R}_i= c(t) \sum_i \mathrm{sgn}(\mu_i) p_i \mathcal{R}_i.
\end{align}
Here, $c(t)=\sum_i |\mu_i|$, $p_i=|\mu_i|/c(t)$, and $\{\mathcal{R}_i\}$ is a set of a polynomial number of physical operations applied for QEM. Each $\mathcal{R}_i$ is a tensor product of single-qubit operations. 
For a given decomposition,
the ideal process $\mathcal{U}_T$ from $t=0$ to $t=T$ can be decomposed as  
\begin{equation}
\begin{aligned}
\mathcal{U}_T&\approx \prod_{n=0}^{N_d-1} \mathcal{E}_I  (n \delta t) \\
&=C(T) \sum_{\vec{i}} p_{\vec{i}} s_{\vec{i}} \prod_{n=0}^{N_d-1} \mathcal{R}_{i_n} \mathcal{E}_N (n \delta t) +O(T \delta t),
\end{aligned}
\end{equation}
where  $N_d=T/\delta t$, $\vec{i}=(i_1, i_2,...,i_{N_d})$, $p_{\vec{i}}=\prod_{n=0}^{N_d-1}p_{i_n}$, $s_{\vec{i}}=\prod_{n=0}^{N_d-1}\mathrm{sgn} (\mu_{i_n})$, and the QEM cost can be described as $C(T)=\prod_{n=0}^{N_d-1} c(n \delta t)$.
Supposing that the initial state for the quantum circuit is
 $\rho_{in}$, we have
\begin{align}
\rho_I (T)= C(T) \sum_{\vec{i}} p_{\vec{i}} s_{\vec{i}} \rho_{\vec{i}} + O(T \delta t),
\end{align}
where $\rho_I(T)$ is the density operator after the ideal process
$\rho_I(T)=\mathcal{U}_T (\rho_{in})$ 
and 
$\rho_{\vec{i}}=(\prod_{n=0}^{N_d-1} \mathcal{R}_{i_n} \mathcal{E}_N (n \delta t))(\rho_{in})$.
%
When measuring an observable $M$, since the expectation value for the state $\rho$ is equal to $\braket{M}_\rho=\mathrm{Tr}[\rho M]$, we obtain
\begin{equation}
\braket{M}_{\rho_I}= C(T) \sum_{\vec{i}} p_{\vec{i}} s_{\vec{i}} \braket{M}_{\rho_{\vec{i}}}+O(T \delta t).
\label{Eq:continuous}
\end{equation}

We can obtain $\braket{M}_{\rho_I}$ of Eq.~(\ref{Eq:continuous}) from the actual experiment
 as follows: First,
 we generate the recovery operation $\mathcal{R}_i$ with a probability
 $p_i$ with a time interval $\delta t$ until time $T$, and measure the observable $M$, and we record the measurement outcome after multiplying the factor of $s_{\vec{i}}$.  
 Second, we repeat the same procedure to reduce the statistical uncertainty. Finally, we estimate the value of $ C(T) \sum_{\vec{i}} p_{\vec{i}} s_{\vec{i}} \braket{M}_{\rho_{\vec{i}}}$ from the measurement results, and this
approximates the error-free expectation value.  

Since $\mathcal{E}_N \approx \mathcal{E}_I$ for a small $\delta t$ and the recovery operation becomes an identity operation in almost all the cases, we can use the Monte Carlo method to stochastically 
realize
continuous recovery operations $\mathcal{R}_i$ corresponding to $\delta t \rightarrow +0$ to eliminate a 
discretization
error $O(T \delta t)$.  This procedure is similar to the one employed in the simulation of the stochastic Schr\"odinger equation (refer to Ref.~\cite{sun2020practical} for details). Letting $c(t)=1+ c'(t) \delta t$, the QEM cost becomes
\begin{equation}
\begin{aligned}
C(T)&= \lim_{\delta t \rightarrow +0} \prod_{n=0}^{T/\delta t -1} (1+c'(n \delta t) \delta t) \\
&= \mathrm{exp}\bigg(\int_0^{T} dt c'(t) \bigg).
\end{aligned}
\end{equation}

\section{Properties of Non-Markovianity}
In this section, we review typical properties of non-Markovianity~\cite{rivas2014quantum,breuer2016colloquium}.
\subsection{Definition of non-Markovianity}
There are some definitions of Markovianity in quantum dynamics, but
throughout this paper, we adopt a definition introduced in Refs~\cite{wolf2008dividing,rivas2010entanglement}.
Here, a Markovian map is defined  as {\it a CP-divisible map }:
a dynamical map $\mathcal{E}_{(t,0)}$ from $0$ to $t$ is CP divisible if the map $\mathcal{E}_{(t,s)}$ $(0\le s\le t)$ defined by
\begin{align}
    \mathcal{E}_{(t,s)}=\mathcal{E}_{(t,0)}\mathcal{E}_{(s,0)}^{-1}
\end{align}
is completely positive for all time $s$.
Otherwise, dynamical maps are non-Markovian.

Particularly when the inverse of dynamical maps $\mathcal{E}_{(t,0)}$ exists, even if the dynamical maps are non-Markovian, the equations of the dynamics can be written in the canonical form of the time-local master equation~\cite{hall2014canonical,de2017dynamics}
\begin{align}
\frac{d \rho(t)}{dt}=&-i[H(t), \rho(t)]+\sum_k \gamma_k(t)\Bigl[L_k(t)\rho(t)L_k^\dagger(t)\notag\\
&-\frac{1}{2}\left\{L_k^\dagger(t)L_k(t),\rho(t)\right\}\Bigr],
\label{eqn:timelocalmaseq}
\end{align}
where $\gamma_k(t)$ is a decay rate and $L_k(t)$ is a time-dependent decoherence operator satisfying $\mathrm{Tr}[L_k(t)]=0$ and $\mathrm{Tr}[L_k^\dagger(t)L_k(t)]=1$.
Here, $\{A,B\}=AB+BA$ denotes anticommutator.
It is worth mentioning that Eq.~(\ref{eqn:timelocalmaseq}) has a similar form to the  time-dependent Lindblad Markovian master equation, except that
the decay rate $\gamma_k(t)$ can be negative in some time interval.
In fact, if and only if all the decay rates $\gamma_k(t)$ are non-negative for all the time $t$, the dynamical maps are CP divisible~\cite{hall2014canonical}.
In other words, the sign of $\gamma_k(t)$ characterizes whether the dynamical maps are Markovian or non-Markovian.
Compared with the Lindblad master equation, the time-local master equation in Eq.~(\ref{eqn:timelocalmaseq}) is a general equation that can describe many cases of non-Markovian noise and the cases of strong coupling between the system and the environments, which cannot be described by the Lindblad master equation under the assumptions of the weak coupling limit.

\subsection{Measure of non-Markovianity}
There are several non-Markovian measures proposed in previous studies (for example, see review papers~\cite{rivas2014quantum,breuer2016colloquium}).
In this paper, to quantify non-Markovianity, we adopt the decay rate measure~\cite{hall2014canonical}:
\begin{align}
F(t',t)&=\sum_k\int_t^{t'}ds  \frac{|\gamma_k(s)|-\gamma_k(s)}{2}.
\label{eqn:nonmarkovmeas}
\end{align}
Since Markovian dynamical maps give $F(t,t')=0$, this measure can be interpreted as the total amount of non-Markovianity.
Moreover, it is shown that this measure is equivalent to the RHP  measure~\cite{rivas2010entanglement,hall2014canonical} except for a constant factor, which quantifies the degree of non-complete positive of the map $\mathcal{E}_{(t,s)}$ based on the Choi–Jamio{\l}kowski isomorphism~\cite{choi1975completely,jamiolkowski1972linear}.

\section{Relation between QEM costs and the measure for non-Markovianity}
We derive QEM costs for non-Markovian dynamics and discuss the direct relation between  costs and the measure of non-Markovianity.
Note that, since the time-local master equation [Eq.~(\ref{eqn:timelocalmaseq})] is derived from a given Hamiltonian, modifications of the Hamiltonian for applying recovery operations could affect the form of time-local master equation. 
However, to derive a relation between QEM costs and non-Markovian measures, we assume that recovery operations 
do not change Eq.~(\ref{eqn:timelocalmaseq}). 
\subsection{General form of QEM costs}
Here, we derive the general form of QEM costs for the time-local quantum master equation,  Eq.~(\ref{eqn:timelocalmaseq}).
The key idea is to represent the decoherence operators $L_k(t)$ using the process matrix form~\cite{greenbaum2015introduction}
\begin{align}
L_k(t)&=\sum_{i= 1}^{d^2-1}  \frac{1}{d^2}\mathrm{Tr}[L_k(t) G_i]G_i,
\end{align}
where the operators $G_i$ satisfy the conditions
$G_0=I^{\otimes N}$, $ G_i=G_i^\dagger$, $ \mathrm{Tr}[G_iG_j]=d\times\delta_{ij}$, and $ (G_i)^2=I^{\otimes N}$, and $d=2^N$ is the dimension of the state vector of $N$ qubits.
An example of $\{G_i\}_i$ is a set of Pauli products, i.e., $G_i\in \{I, X, Y, Z\}^{\otimes N}$.
Then, Eq.~(\ref{eqn:timelocalmaseq}) can be rewritten as
\begin{align}
\frac{d}{d t}\rho_N(t)&=\sum_{i,j= 0}^{d^2-1} M_{ij}(t)G_i\rho_N(t)G_j,
\end{align}
where $M(t)$ is a $d^2\times d^2$ Hermitian matrix defined by
\begin{align}
M_{ij}(t)= \begin{cases}
    \frac{1}{d^4}\sum_k \gamma_k(t)\mathrm{Tr}[L_k(t) G_i]\mathrm{Tr}[L_k^\dagger(t) G_j] & (i,j\ge1) \\
    -\frac{1}{2d^2}\sum_k \gamma_k(t)\mathrm{Tr}[L_k^\dagger(t)L_k(t) G_i] & (i\ge1, j=0)\\
    -\frac{1}{2d^2}\sum_k \gamma_k(t)\mathrm{Tr}[L_k^\dagger(t)L_k(t) G_j] & (i=0, j\ge1)\\
    -\sum_k \gamma_k(t) & (i=j=0)
  \end{cases}\notag
\end{align}
and $M(t)$ can be diagonalized using a unitary matrix $u$:
\begin{align}
M_{ij}(t)&=\sum_{l=0}^{d^2-1} u_{il}(t)q_l(t) u_{jl}^*(t).
\end{align}
Therefore, we obtain
\begin{align}
\frac{d}{d t}\rho_N(t)
&=\sum_{l=0}^{d^2-1} q_l(t) B_l(t) \rho_N(t) B_l^\dagger(t),
\label{eqn:generalformoperation}
\end{align}
where $B_l(t)$ is an operator $B_l(t)=\sum_{i=0}^{d^2-1}u_{il}(t)G_i$.

Using Eq.~(\ref{eqn:generalformoperation}), we can derive the QEM cost.
By choosing the recovery operation at time $t$,
\begin{align}
\mathcal{E}_Q (t)&=c(t)\left(p_0(t) \mathcal{I}- \sum_{l\ge 1} {\rm sgn} \left(q_l(t)\right)p_l(t)\mathcal{B}_l(t)\right),
\end{align}
where $c(t)=1+[-q_0(t)+\sum_{l\ge 1} |q_l(t)|]\delta t$, $p_l(t)= |q_l(t)|\delta t$ ($l\ge 1$), $p_0(t)= 1-\sum_{l\ge 1} p_l(t)$, and $\mathcal{B}_l(t) \rho=B_l(t)\rho B_l^\dagger(t)$.
Hence, the general form of QEM costs is given by
\begin{align}
C(T)&=\exp{\left[\int_0^T\left(-q_0(t)+\sum_{l\ge 1} |q_l(t)|\right)dt\right]}.
\end{align}

\subsection{The effect of non-Markovianity on QEM costs}
We further assume $L_k^\dagger(t)L_k(t)=L_k(t)L_k^\dagger(t)=I^{\otimes N}$ for all $k$ in Eq.~(\ref{eqn:timelocalmaseq}), such as Pauli products.
In this case, the matrix $M$ can be easily diagonalized because $M_{i0}=0$ and $M_{0i}=0$ for all $i$ ($i\ge 1$) and the unitary matrix $u$ is determined by $u_{ik}=\frac{1}{d^2}\mathrm{Tr}[L_k(t) G_i]$ $(i,k\ge 1)$, $u_{i0}=u_{0l}=0$ $(i,l\ge 1)$, and $u_{00}=1$, and the eigenvalues of $M$ are $q_k=\gamma_k(t)$ $(k\ge 1)$ and $q_0=-\sum_{k\ge 1}\gamma_k(t)$, and the recovery operations are $\mathcal{B}_k(t) \rho=L_k(t)\rho L_k^\dagger(t)$ $(k\ge 1)$.
We can derive the QEM costs as
\begin{align}
C(T)=&\exp{\left[\sum_k\int_0^{T}  \Bigl(|\gamma_k(t)|+\gamma_k(t)\Bigr)dt\right]}.
\label{eqn:QEMcost}
\end{align}
Here, we define the quantity $D(t',t)=\sum_k\int_t^{t'}ds  |\gamma_k(s)|$, which is equivalent to the QEM cost $e^{D(t',t)}$ for the Markovian case with decay rates $|\gamma_k(t)|$.
By using $D(t',t)$ and $F(t',t)$, 
we can rewrite the QEM cost as
\begin{align}
C(T)=&\exp{\left[2\Bigl(D(T,0)-F(T,0)\Bigr)\right]}.
\end{align}
From this equation, we can understand that as the amount of non-Markovianity in Eq.~(\ref{eqn:nonmarkovmeas}) increases, the QEM cost is reduced. 
More specifically, in a time region with  $\gamma_k(t)<0$ for all $k$, the QEM cost does not increase at all.


\subsection{Study of specific models}
Here, we study the QEM cost for specific models. Although the implementation of recovery
operations could change the form of the time-local master equation, 
 we discuss the case in which it is invariant; i.e., recovery operations commute with both the unitary dynamics of the system Hamiltonian and the noise operators. In this case, since we can perform 
 the recovery operations at the end of the dynamics, the time-local master equation is not affected by recovery processes.

We consider a two-qubit system where a long-lived qubit is coupled with a short-lived qubit. (Another example of
a qubit dispersively coupled with a dissipative resonator 
is illustrated in the Appendix.)
Importantly, this system has been 
realized
with nuclear magnetic resonance, and the non-Markovian noise has been controlled by implementation of the pulse~\cite{matsuzaki2019realization,Kukita_2020}.
The equation of the two-qubit model is given by 
\begin{align}
\frac{d \rho_N^{(1+2)}(t)}{dt}=&i\left[ \rho_N^{(1+2)}(t),\frac{J}{4}Z\otimes Z\right]+ \mathcal{L}\big[\rho_N^{(1+2)}(t)\big], \notag\\
\mathcal{L}\big[\rho\big]=&\sum_{k=1}^2\frac{\gamma_k}{4}\Bigl(2L_k\rho L_k^\dagger-\{L_k^\dagger L_k,\rho\}\Bigr),
\label{eqn:nonmarkovianmaseq}
\end{align}
where $\rho_N^{(1+2)}$ denotes the density operator of the two-qubit system,
$J$ denotes a coupling strength between the qubits, 
$\gamma_1=2\gamma s$ denotes an energy relaxation rate associated with 
a Lindblad operator $L_1=I\otimes \sigma_+$, 
$\gamma_2=2\gamma (1-s)$ denotes an energy relaxation rate associated with 
a Lindblad operator $L_2=I\otimes \sigma_-$, $\sigma_+=(X+ iY)/2$ [$\sigma_-=(X- iY)/2$] denotes a raising (lowering) operator, and 
$s$ denotes a control parameter determined by the environmental temperature
($0\le s \le 1/2$).
Only the second qubit is directly coupled to the Markovian environment and the first qubit is influenced by the environment through the second qubit.
Here,  $1/J$ denotes a time scale of the exchange of the information between the first qubit and the second qubit, while $1/\gamma$ denotes the decoherence time for the second qubit. In the regime of  $J/\gamma>1$ where the information exchange between qubits occurs in a faster time scale than the environmental decoherence of the second qubit, 
 the dynamics of the reduced density operator of the first qubit could be non-Markovian.
This means that the first qubit receives the backflow of the information from the second qubit 
before 
losing the coherent information to the environment.

Here, we choose an initial state as $\rho_N^{(1+2)}(0)=\ket{+}\bra{+}\otimes \rho_{{\rm Gibbs}}$, where $\ket{+}=(\ket{0}+\ket{1})/2$ is the superposition of the computational basis $\ket{0}$ and $\ket{1}$, and $\rho_{{\rm Gibbs}}=s\ket{0}\bra{0}+(1-s)\ket{1}\bra{1}$ is the Gibbs state corresponding to the Lindbladian in Eq.~(\ref{eqn:nonmarkovianmaseq}). 
In this case, the equation of its dynamics is given by
\begin{align}
\frac{d \rho_N^{(1)}(t)}{dt}=&i\left[ \rho_N^{(1)}(t),\frac{S(t)}{2}Z\right] \notag\\
&+\frac{\gamma(t)}{2}\Bigl( Z\rho_N^{(1)}(t)Z -\rho_N^{(1)}(t)\Bigr),
\label{eqn:timelocalmaseqexample1}
\end{align}
where $\rho_N^{(1)}(t)$ is the reduced density operator of the first qubit and $Z$ is a Pauli $Z$ matrix.
In this case, the decay rate $\gamma(t)$ and $S(t)$ are given by 
\begin{align}
\gamma(t)-iS(t)=& -\frac{1}{f(t)}\frac{d}{dt}f(t),
\label{eqn:decayrateex}
\end{align}
where $f(t)=\frac{iJ(2s-1)}{2}\frac{e^{\lambda_+t}-e^{\lambda_-t}}{\lambda_+-\lambda_-}-\frac{\lambda_- e^{\lambda_+t}-\lambda_+ e^{\lambda_-t}}{\lambda_+-\lambda_-}$ and $\lambda_{\pm}$ are the two solutions of an equation $\lambda^2 +\gamma\lambda +[2iJ\gamma (1-2s)+ J^2]/4=0$.
From Eq.~(\ref{eqn:QEMcost}), the QEM cost $C(T)$ can be derived as 
\begin{align}
C(T)=\exp{\left(\int_0^{T}dt  \frac{|\gamma(t)|+\gamma(t)}{2}\right)}
\label{eqn:costdephasing}
\end{align}
by choosing the recovery operation $\mathcal{E}_Q (t)=\left(1+\delta t\frac{\gamma(t)+|\gamma(t)|}{2}\right)\Bigl[\left(1-\delta t\frac{|\gamma(t)|}{2}\right) \mathcal{I}-\mathrm{sgn}\left(\gamma(t)\right)\delta t\frac{|\gamma(t)|}{2}\mathcal{Z}\Bigr]$, where $\mathcal{Z}$ is the operation $\mathcal{Z}\rho=Z\rho Z$.

\begin{figure}
  \includegraphics[width=0.5\textwidth]{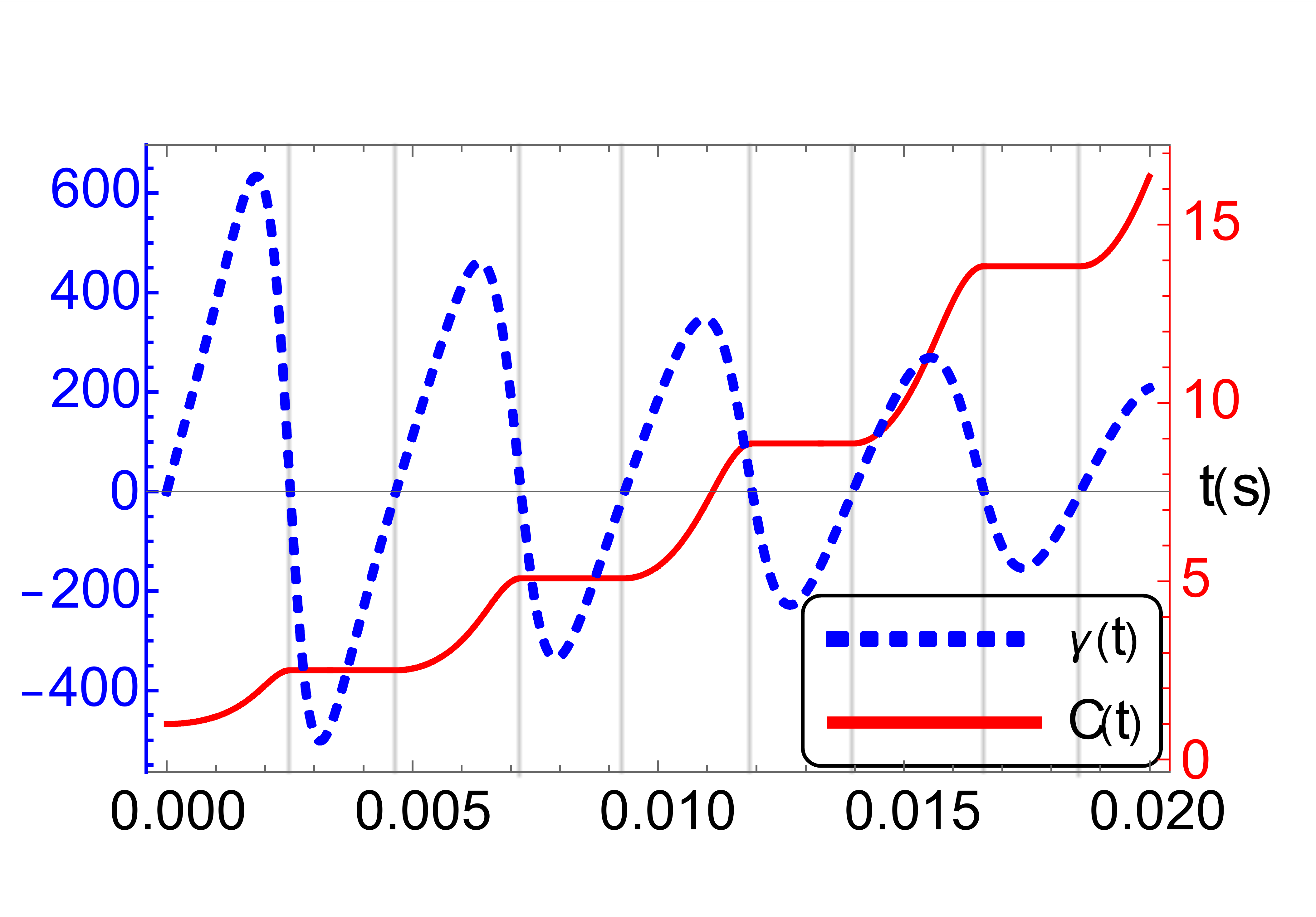}
  \caption{(Color online) The blue dashed line shows the decay rate $\gamma(t)$ s$^{-1}$ in Eq.~(\ref{eqn:decayrateex}) and the red solid line shows the QEM cost $C(t)$ in Eq.~(\ref{eqn:costdephasing}) at time $t$ s.
  We choose the parameters as $J=2\pi\times 215$ rad s$^{-1}$, $\gamma=6.5$ ms with experimental parameters in Ref.~\cite{matsuzaki2019realization}, and moreover the parameter of the finite-temperature effect as $s=0.3$. The vertical lines represent the solutions satisfying $\gamma(t)=0$.}
  \label{fig:gammacostNMR}
\end{figure}
Figure \ref{fig:gammacostNMR} shows the results of $\gamma(t)$ and $C(t)$ at time $t$ with experimental parameters in Ref.~\cite{matsuzaki2019realization}. 
In Fig.~\ref{fig:gammacostNMR}, $\gamma(t)$ becomes negative in some time interval and therefore the dynamics is actually non-Markovian. 
Moreover, the area where $\gamma(t)\ge 0$ holds only contributes to the QEM cost $C(t)$, as shown in Fig.~\ref{fig:gammacostNMR}. 
In other words, the QEM cost $C(t)$ does not increase at all in the region satisfying $\gamma(t)< 0$, and the area of its region is equivalent to the non-Markovian measure in Eq.~(\ref{eqn:nonmarkovmeas}).

To conclude this section, we note the limit of the application of the time-local master equation in Eq.~(\ref{eqn:timelocalmaseq}) by using this specific model.
We set the temperature of the environment to infinity $(s=1/2)$. In this case, we cannot write the dynamics as time-local form because $\gamma(t)$ diverges at a specific time point. However, this is an artificial effect because the temperature of the realistic system is finite. We can exclude this artificial case, which is not an interesting case in practice.

\section{Discussions}
In this work, we discuss a relationship between the QEM and non-Markovian measures.
Non-Markovianity is characterized by a negative decay rate of the dissipator.
Interestingly, the QEM cost does not increase at all when all decay rates are negative.
This demonstrates that non-Markovianity can contribute to reduce the cost of the QEM.
We show specific physical systems as examples that support our theoretical analysis.

We provide an intuitive explanation for this below: QEM mitigates physical errors which can be attributed to quantum information escaping to the environment, and in the case of Markovian dynamics, such information never comes back to the system. Therefore, the QEM cost monotonously increases. However, in the case of non-Markovian dynamics, the quantum information can return to the system from the environment, which contributes to the reduction of the sampling cost of QEM.

Furthermore, we mention the difference between QEM for non-Markovian cases and dynamical decoupling.
Dynamical decoupling is a method to cancel the interaction between the system and the environment for non-Markovian cases, and we can in principle obtain a noiseless state. 
On the other hand, QEM is the method to obtain just an expectation value of a noiseless state. Importantly, QEM does not recover the quantum state itself, which is a significant difference from the dynamical decoupling. 

We focus on the case where the decoherence operators can be described by a set of orthogonal operators such as Pauli operators, and leave more general cases for a future work. 
Moreover, the intentional introduction of the non-Markovianity might  contribute to the QEM. 
For this purpose, we need to control not the system but the environment.
Actually, there are some theoretical proposals and experimental demonstrations about the control of the non-Markovianity \cite{Kukita_2020,PhysRevA.98.053862,Valente:16}, which might lead to the potential use of the non-Markovianity to improve the efficiency of the QEM. It is worth noting that the intentional introduction of the non-Markovianity is different from dynamical decoupling, because dynamical decoupling is not a control of the environment but a control of the system.
Our work helps in understanding the properties of QEM and may lead to sophisticated construction of QEM for realistic quantum systems with non-Markovian noise.

\begin{acknowledgments}
This work was supported by the Leading Initiative for Excellent Young Researchers MEXT Japan and JST presto (Grant No.~JPMJPR1919), Japan.
This paper is partly based on results obtained from a project, JPNP16007, commissioned by the New Energy and Industrial Technology Development Organization (NEDO), Japan. This work was supported by MEXT Quantum Leap
Flagship Program (MEXT Q- LEAP) (Grants No.~JPMXS0120319794, and No.~JPMXS0118068682) and JST ERATO (Grant No.~JPMJER1601).
\end{acknowledgments}

\appendix

\section{A qubit dispersively coupled with a dissipative resonator}
Here, we also consider a qubit-resonator system where a qubit is dispersively coupled with a lossy resonator. The qubit is affected by a dephasing induced from the interaction with the resonator, and this dynamics has been studied in Refs.~\cite{govenius2015parity,blais2004cavity,bertet2005dephasing}. 
When the frequency of the qubit is significantly detuned from that of the resonator, the Hamiltonian is given by

\begin{align}
H_{q+r}=&\chi Z a^\dagger a,
\label{eqn:dispersiveHamiltonian}
\end{align}
where $\chi/\pi$ is the dispersive frequency shift of the qubit per photon and $a^\dagger$ ($a$) is the creation (annihilation) operator of the photon in the resonator.
For simplicity, we assume that we are in a rotating frame, and we only consider the interaction Hamiltonian in Eq.~(\ref{eqn:dispersiveHamiltonian}).
The dynamics of the system can be described by the Lindblad master equation
\begin{align}
\frac{d \rho_N^{(q+r)}(t)}{dt}=&i\left[ \rho_N^{(q+r)}(t),H_{q+r}\right] \notag\\
+&\frac{\kappa}{2}\Bigl( 2a\rho_N^{(q+r)}(t)a^\dagger -\{a^\dagger a,\rho_N^{(q+r)}(t)\}\Bigr),
\label{eqn:dispersivenonmarkovianmaseq}
\end{align}
where $\kappa$ is a decay rate.
We set the initial state to $\ket{+}\otimes \ket{\alpha}$, where 
$\ket{\alpha}$ is a coherent state for $\alpha \in \mathbb{C}$.
This equation can be easily solved as $\rho_N^{(q+r)}(t)=\sum_{i,j=0}^1 c_{ij}(t)\ket{i}\bra{j}\otimes \ket{\alpha_i(t)}\bra{\alpha_j(t)}$, where $c_{00}(t)=c_{11}(t)=\frac{1}{2}$ and $c_{10}(t)=c_{01}(t)^*$ and
\begin{align}
c_{01}(t)=\frac{c_{01}(0)}{\braket{\alpha_1(t)|\alpha_0(t)}}\exp{\left[-|\alpha|^2\frac{1-e^{(2i\chi-\kappa)t}}{1-i\kappa/2\chi}\right]}.
\end{align}
Here, $\alpha_0(t)=e^{(i\chi-\kappa/2)t}\alpha$ and $\alpha_1(t)=e^{(-i\chi-\kappa/2)t}\alpha$.
The reduced density operator of the qubit is given by $\rho_N^{(q)}(t)=\sum_{i,j=0}^1 c'_{ij}(t)\ket{i}\bra{j}$, where $c'_{00}(t)=c'_{11}(t)=\frac{1}{2}$ and $c'_{10}(t)=c'_{01}(t)^*$ and $c'_{01}(t)=c_{01}(t)\times e^{(2i\chi-\kappa)t}=\frac{1}{2}e^{-|\alpha|^2(x+iy)}$.
From this, the time-local master equation can be derived as
\begin{align}
\frac{d \rho_N^{(q)}(t)}{dt}=&i\left[ \rho_N^{(q)}(t),\frac{S(t)}{2}Z\right] \notag\\
+&\frac{\gamma(t)}{2}\Bigl( Z\rho_N^{(q)}(t)Z -\rho_N^{(q)}(t)\Bigr),
\label{eqn:dispersivenonmarkoviantimelocalmaseq}
\end{align}
where $S(t)=|\alpha|^2\frac{dx}{dt}$ and the decay rate of the qubit,  $\gamma(t)=|\alpha|^2\frac{dy}{dt}$, are given by
\begin{align}
S(t)=&|\alpha|^2 e^{-\kappa t}(\kappa(1-\cos{2\chi t})-2\chi \sin{2\chi t})\notag\\
&+ \frac{|\alpha|^2e^{-\kappa t}}{1+(\kappa/2\chi)^2}\left(2\kappa \cos{2\chi t}+\left(2\chi - \frac{\kappa^2}{2\chi}\right)\sin{2\chi t}\right),\\
\gamma(t)=&\frac{|\alpha|^2\kappa e^{-\kappa t}}{1+(\kappa/2\chi)^2}\left(\frac{\kappa}{\chi} \cos{2\chi t}+\left(1-\left(\frac{\kappa}{2\chi}\right)^2\right)\sin{2\chi t}\right).
\label{eqn:decayrateex2}
\end{align}
From Eq.~(\ref{eqn:decayrateex2}), $\gamma(t)$ behaves as a damped oscillation with the time constant $\kappa$ and the angular frequency $2\chi$.
The QEM cost $C(T)$ is the same form as that in Sec.~IV C 
and therefore $C(T)$ can decrease in the case of $\kappa < \chi$. 

\begin{figure}[h!]
  \includegraphics[width=0.45\textwidth]{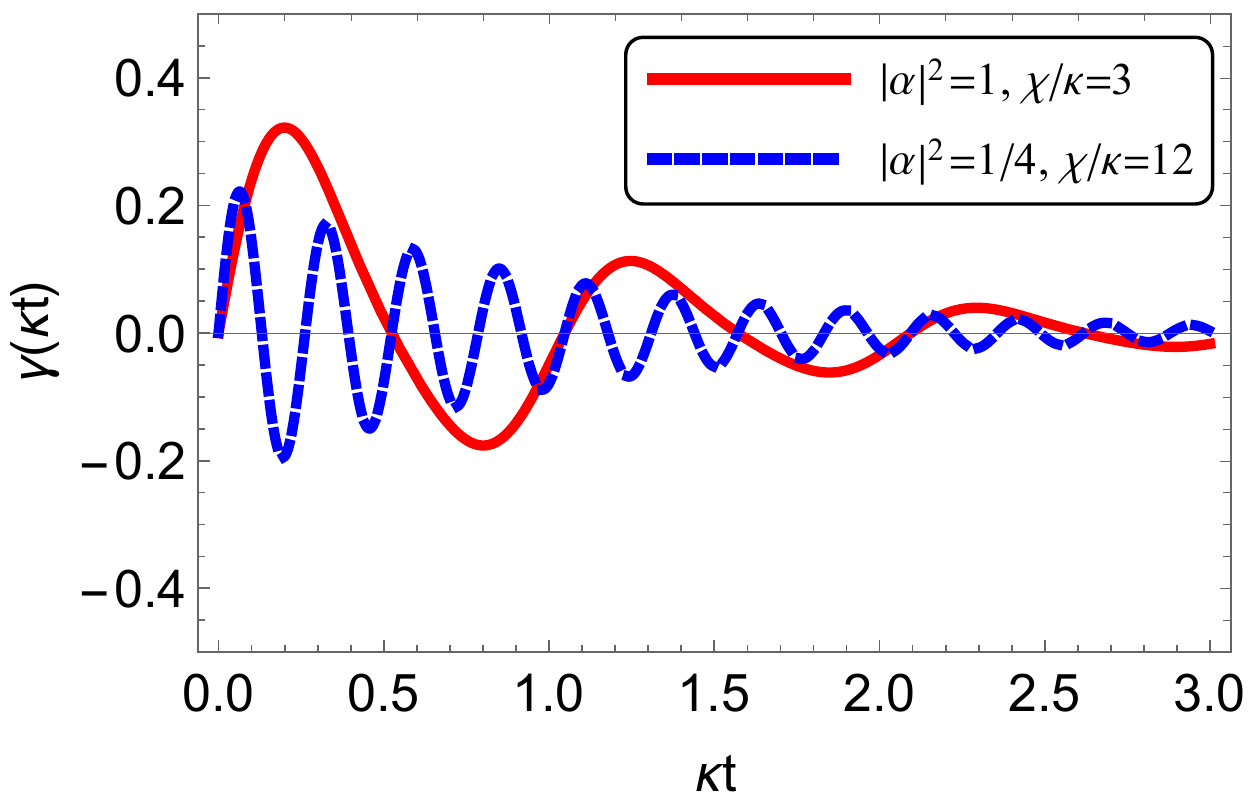}
  \caption{(Color online) The decay rate $\gamma(\kappa t)$ in Eq.~(\ref{eqn:decayrateex2}) at $\kappa t$.
  We choose the parameters as $|\alpha|^2=1,\chi/\kappa=3$ (red solid line) and $|\alpha|^2=1/4,\chi/\kappa=12$ (blue dashed line).}
  \label{fig:gammadispersivecoupling}
  \includegraphics[width=0.45\textwidth]{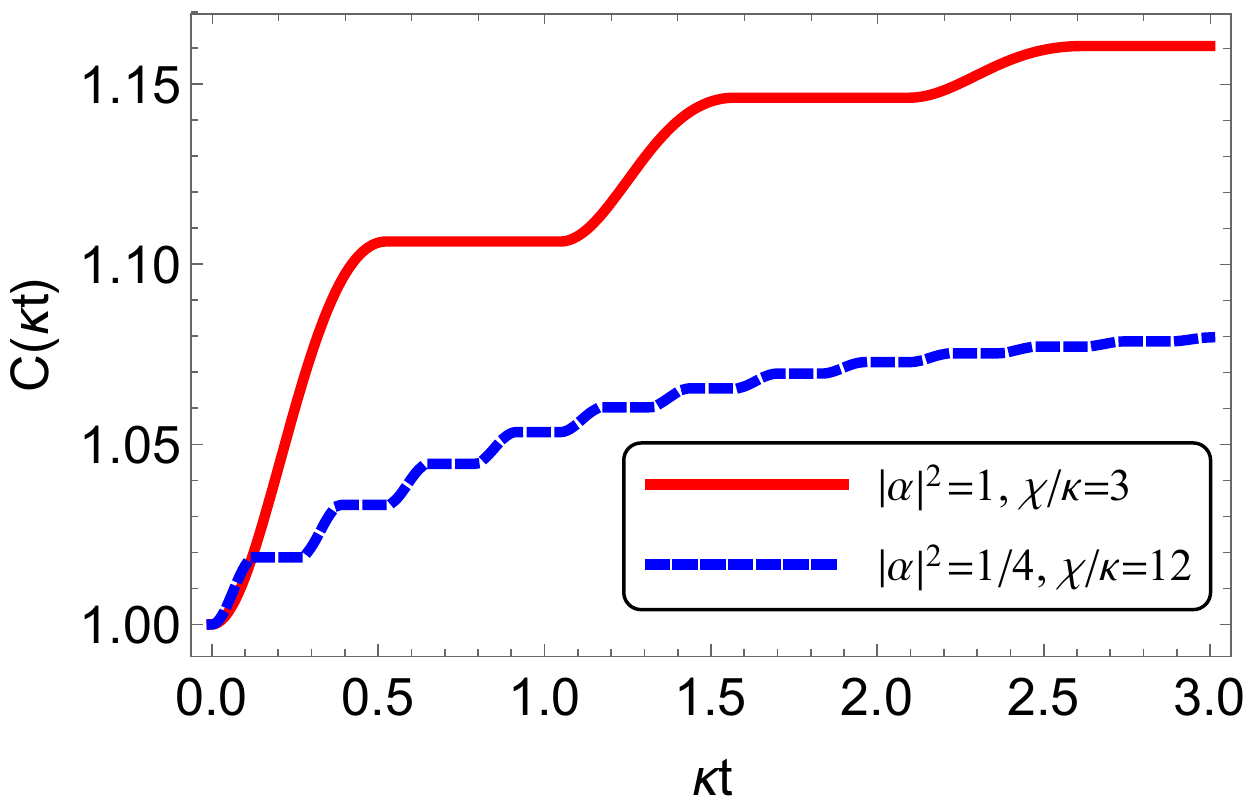}
  \caption{(Color online) The QEM cost $C(\kappa t)$ at $\kappa t$. All the parameters are the same as those in Fig.~\ref{fig:gammadispersivecoupling}.}
  \label{fig:costdispersivecoupling}
\end{figure}
In Figs.~\ref{fig:gammadispersivecoupling} and \ref{fig:costdispersivecoupling}, we show the numerical results of $\gamma(\kappa t)$ and $C(\kappa t)$. 
As is the same in Sec.~IV C, the dynamics is actually non-Markovian because of some negative regions of $\gamma(\kappa t)$ in Fig.~\ref{fig:gammadispersivecoupling}, and the QEM cost $C(\kappa t)$ do not increase at all for those regions.

\bibliographystyle{apsrev4-1}
\bibliography{nonmarkovmitigation}

\begin{thebibliography}{48}%
\makeatletter
\providecommand \@ifxundefined [1]{%
 \@ifx{#1\undefined}
}%
\providecommand \@ifnum [1]{%
 \ifnum #1\expandafter \@firstoftwo
 \else \expandafter \@secondoftwo
 \fi
}%
\providecommand \@ifx [1]{%
 \ifx #1\expandafter \@firstoftwo
 \else \expandafter \@secondoftwo
 \fi
}%
\providecommand \natexlab [1]{#1}%
\providecommand \enquote  [1]{``#1''}%
\providecommand \bibnamefont  [1]{#1}%
\providecommand \bibfnamefont [1]{#1}%
\providecommand \citenamefont [1]{#1}%
\providecommand \href@noop [0]{\@secondoftwo}%
\providecommand \href [0]{\begingroup \@sanitize@url \@href}%
\providecommand \@href[1]{\@@startlink{#1}\@@href}%
\providecommand \@@href[1]{\endgroup#1\@@endlink}%
\providecommand \@sanitize@url [0]{\catcode `\\12\catcode `\$12\catcode
  `\&12\catcode `\#12\catcode `\^12\catcode `\_12\catcode `\%12\relax}%
\providecommand \@@startlink[1]{}%
\providecommand \@@endlink[0]{}%
\providecommand \url  [0]{\begingroup\@sanitize@url \@url }%
\providecommand \@url [1]{\endgroup\@href {#1}{\urlprefix }}%
\providecommand \urlprefix  [0]{URL }%
\providecommand \Eprint [0]{\href }%
\providecommand \doibase [0]{http://dx.doi.org/}%
\providecommand \selectlanguage [0]{\@gobble}%
\providecommand \bibinfo  [0]{\@secondoftwo}%
\providecommand \bibfield  [0]{\@secondoftwo}%
\providecommand \translation [1]{[#1]}%
\providecommand \BibitemOpen [0]{}%
\providecommand \bibitemStop [0]{}%
\providecommand \bibitemNoStop [0]{.\EOS\space}%
\providecommand \EOS [0]{\spacefactor3000\relax}%
\providecommand \BibitemShut  [1]{\csname bibitem#1\endcsname}%
\let\auto@bib@innerbib\@empty
\bibitem [{\citenamefont {Nielsen}\ and\ \citenamefont
  {Chuang}(2002)}]{nielsen2002quantum}%
  \BibitemOpen
  \bibfield  {author} {\bibinfo {author} {\bibfnamefont {M.~A.}\ \bibnamefont
  {Nielsen}}\ and\ \bibinfo {author} {\bibfnamefont {I.}~\bibnamefont
  {Chuang}},\ }\href@noop {} {\enquote {\bibinfo {title} {Quantum computation
  and quantum information},}\ } (\bibinfo {year} {2002})\BibitemShut {NoStop}%
\bibitem [{\citenamefont {Georgescu}\ \emph {et~al.}(2014)\citenamefont
  {Georgescu}, \citenamefont {Ashhab},\ and\ \citenamefont
  {Nori}}]{georgescu2014quantum}%
  \BibitemOpen
  \bibfield  {author} {\bibinfo {author} {\bibfnamefont {I.~M.}\ \bibnamefont
  {Georgescu}}, \bibinfo {author} {\bibfnamefont {S.}~\bibnamefont {Ashhab}}, \
  and\ \bibinfo {author} {\bibfnamefont {F.}~\bibnamefont {Nori}},\ }\href@noop
  {} {\bibfield  {journal} {\bibinfo  {journal} {Reviews of Modern Physics}\
  }\textbf {\bibinfo {volume} {86}},\ \bibinfo {pages} {153} (\bibinfo {year}
  {2014})}\BibitemShut {NoStop}%
\bibitem [{\citenamefont {Harrow}\ \emph {et~al.}(2009)\citenamefont {Harrow},
  \citenamefont {Hassidim},\ and\ \citenamefont {Lloyd}}]{harrow2009quantum}%
  \BibitemOpen
  \bibfield  {author} {\bibinfo {author} {\bibfnamefont {A.~W.}\ \bibnamefont
  {Harrow}}, \bibinfo {author} {\bibfnamefont {A.}~\bibnamefont {Hassidim}}, \
  and\ \bibinfo {author} {\bibfnamefont {S.}~\bibnamefont {Lloyd}},\
  }\href@noop {} {\bibfield  {journal} {\bibinfo  {journal} {Physical review
  letters}\ }\textbf {\bibinfo {volume} {103}},\ \bibinfo {pages} {150502}
  (\bibinfo {year} {2009})}\BibitemShut {NoStop}%
\bibitem [{\citenamefont {Hauke}\ \emph {et~al.}(2012)\citenamefont {Hauke},
  \citenamefont {Cucchietti}, \citenamefont {Tagliacozzo}, \citenamefont
  {Deutsch},\ and\ \citenamefont {Lewenstein}}]{hauke2012can}%
  \BibitemOpen
  \bibfield  {author} {\bibinfo {author} {\bibfnamefont {P.}~\bibnamefont
  {Hauke}}, \bibinfo {author} {\bibfnamefont {F.~M.}\ \bibnamefont
  {Cucchietti}}, \bibinfo {author} {\bibfnamefont {L.}~\bibnamefont
  {Tagliacozzo}}, \bibinfo {author} {\bibfnamefont {I.}~\bibnamefont
  {Deutsch}}, \ and\ \bibinfo {author} {\bibfnamefont {M.}~\bibnamefont
  {Lewenstein}},\ }\href@noop {} {\bibfield  {journal} {\bibinfo  {journal}
  {Reports on Progress in Physics}\ }\textbf {\bibinfo {volume} {75}},\
  \bibinfo {pages} {082401} (\bibinfo {year} {2012})}\BibitemShut {NoStop}%
\bibitem [{\citenamefont {Bylander}\ \emph {et~al.}(2011)\citenamefont
  {Bylander}, \citenamefont {Gustavsson}, \citenamefont {Yan}, \citenamefont
  {Yoshihara}, \citenamefont {Harrabi}, \citenamefont {Fitch}, \citenamefont
  {Cory}, \citenamefont {Nakamura}, \citenamefont {Tsai},\ and\ \citenamefont
  {Oliver}}]{bylander2011noise}%
  \BibitemOpen
  \bibfield  {author} {\bibinfo {author} {\bibfnamefont {J.}~\bibnamefont
  {Bylander}}, \bibinfo {author} {\bibfnamefont {S.}~\bibnamefont
  {Gustavsson}}, \bibinfo {author} {\bibfnamefont {F.}~\bibnamefont {Yan}},
  \bibinfo {author} {\bibfnamefont {F.}~\bibnamefont {Yoshihara}}, \bibinfo
  {author} {\bibfnamefont {K.}~\bibnamefont {Harrabi}}, \bibinfo {author}
  {\bibfnamefont {G.}~\bibnamefont {Fitch}}, \bibinfo {author} {\bibfnamefont
  {D.~G.}\ \bibnamefont {Cory}}, \bibinfo {author} {\bibfnamefont
  {Y.}~\bibnamefont {Nakamura}}, \bibinfo {author} {\bibfnamefont {J.-S.}\
  \bibnamefont {Tsai}}, \ and\ \bibinfo {author} {\bibfnamefont {W.~D.}\
  \bibnamefont {Oliver}},\ }\href@noop {} {\bibfield  {journal} {\bibinfo
  {journal} {Nature Physics}\ }\textbf {\bibinfo {volume} {7}},\ \bibinfo
  {pages} {565} (\bibinfo {year} {2011})}\BibitemShut {NoStop}%
\bibitem [{\citenamefont {Reiner}\ \emph {et~al.}(2019)\citenamefont {Reiner},
  \citenamefont {Wilhelm-Mauch}, \citenamefont {Sch{\"o}n},\ and\ \citenamefont
  {Marthaler}}]{reiner2019finding}%
  \BibitemOpen
  \bibfield  {author} {\bibinfo {author} {\bibfnamefont {J.-M.}\ \bibnamefont
  {Reiner}}, \bibinfo {author} {\bibfnamefont {F.}~\bibnamefont
  {Wilhelm-Mauch}}, \bibinfo {author} {\bibfnamefont {G.}~\bibnamefont
  {Sch{\"o}n}}, \ and\ \bibinfo {author} {\bibfnamefont {M.}~\bibnamefont
  {Marthaler}},\ }\href@noop {} {\bibfield  {journal} {\bibinfo  {journal}
  {Quantum Science and Technology}\ }\textbf {\bibinfo {volume} {4}},\ \bibinfo
  {pages} {035005} (\bibinfo {year} {2019})}\BibitemShut {NoStop}%
\bibitem [{\citenamefont {Ma}\ \emph {et~al.}(2019)\citenamefont {Ma},
  \citenamefont {Saxberg}, \citenamefont {Owens}, \citenamefont {Leung},
  \citenamefont {Lu}, \citenamefont {Simon},\ and\ \citenamefont
  {Schuster}}]{ma2019dissipatively}%
  \BibitemOpen
  \bibfield  {author} {\bibinfo {author} {\bibfnamefont {R.}~\bibnamefont
  {Ma}}, \bibinfo {author} {\bibfnamefont {B.}~\bibnamefont {Saxberg}},
  \bibinfo {author} {\bibfnamefont {C.}~\bibnamefont {Owens}}, \bibinfo
  {author} {\bibfnamefont {N.}~\bibnamefont {Leung}}, \bibinfo {author}
  {\bibfnamefont {Y.}~\bibnamefont {Lu}}, \bibinfo {author} {\bibfnamefont
  {J.}~\bibnamefont {Simon}}, \ and\ \bibinfo {author} {\bibfnamefont {D.~I.}\
  \bibnamefont {Schuster}},\ }\href@noop {} {\bibfield  {journal} {\bibinfo
  {journal} {Nature}\ }\textbf {\bibinfo {volume} {566}},\ \bibinfo {pages}
  {51} (\bibinfo {year} {2019})}\BibitemShut {NoStop}%
\bibitem [{\citenamefont {Preskill}(2018)}]{preskill2018quantum}%
  \BibitemOpen
  \bibfield  {author} {\bibinfo {author} {\bibfnamefont {J.}~\bibnamefont
  {Preskill}},\ }\href@noop {} {\bibfield  {journal} {\bibinfo  {journal}
  {Quantum}\ }\textbf {\bibinfo {volume} {2}},\ \bibinfo {pages} {79} (\bibinfo
  {year} {2018})}\BibitemShut {NoStop}%
\bibitem [{\citenamefont {Temme}\ \emph {et~al.}(2017)\citenamefont {Temme},
  \citenamefont {Bravyi},\ and\ \citenamefont {Gambetta}}]{temme2017error}%
  \BibitemOpen
  \bibfield  {author} {\bibinfo {author} {\bibfnamefont {K.}~\bibnamefont
  {Temme}}, \bibinfo {author} {\bibfnamefont {S.}~\bibnamefont {Bravyi}}, \
  and\ \bibinfo {author} {\bibfnamefont {J.~M.}\ \bibnamefont {Gambetta}},\
  }\href@noop {} {\bibfield  {journal} {\bibinfo  {journal} {Physical review
  letters}\ }\textbf {\bibinfo {volume} {119}},\ \bibinfo {pages} {180509}
  (\bibinfo {year} {2017})}\BibitemShut {NoStop}%
\bibitem [{\citenamefont {Endo}\ \emph {et~al.}(2018)\citenamefont {Endo},
  \citenamefont {Benjamin},\ and\ \citenamefont {Li}}]{endo2018practical}%
  \BibitemOpen
  \bibfield  {author} {\bibinfo {author} {\bibfnamefont {S.}~\bibnamefont
  {Endo}}, \bibinfo {author} {\bibfnamefont {S.~C.}\ \bibnamefont {Benjamin}},
  \ and\ \bibinfo {author} {\bibfnamefont {Y.}~\bibnamefont {Li}},\ }\href@noop
  {} {\bibfield  {journal} {\bibinfo  {journal} {Physical Review X}\ }\textbf
  {\bibinfo {volume} {8}},\ \bibinfo {pages} {031027} (\bibinfo {year}
  {2018})}\BibitemShut {NoStop}%
\bibitem [{\citenamefont {Li}\ and\ \citenamefont
  {Benjamin}(2017)}]{li2017efficient}%
  \BibitemOpen
  \bibfield  {author} {\bibinfo {author} {\bibfnamefont {Y.}~\bibnamefont
  {Li}}\ and\ \bibinfo {author} {\bibfnamefont {S.~C.}\ \bibnamefont
  {Benjamin}},\ }\href@noop {} {\bibfield  {journal} {\bibinfo  {journal}
  {Physical Review X}\ }\textbf {\bibinfo {volume} {7}},\ \bibinfo {pages}
  {021050} (\bibinfo {year} {2017})}\BibitemShut {NoStop}%
\bibitem [{\citenamefont {McArdle}\ \emph {et~al.}(2020)\citenamefont
  {McArdle}, \citenamefont {Endo}, \citenamefont {Aspuru-Guzik}, \citenamefont
  {Benjamin},\ and\ \citenamefont {Yuan}}]{mcardle2020quantum}%
  \BibitemOpen
  \bibfield  {author} {\bibinfo {author} {\bibfnamefont {S.}~\bibnamefont
  {McArdle}}, \bibinfo {author} {\bibfnamefont {S.}~\bibnamefont {Endo}},
  \bibinfo {author} {\bibfnamefont {A.}~\bibnamefont {Aspuru-Guzik}}, \bibinfo
  {author} {\bibfnamefont {S.~C.}\ \bibnamefont {Benjamin}}, \ and\ \bibinfo
  {author} {\bibfnamefont {X.}~\bibnamefont {Yuan}},\ }\href@noop {} {\bibfield
   {journal} {\bibinfo  {journal} {Reviews of Modern Physics}\ }\textbf
  {\bibinfo {volume} {92}},\ \bibinfo {pages} {015003} (\bibinfo {year}
  {2020})}\BibitemShut {NoStop}%
\bibitem [{\citenamefont {Sun}\ \emph {et~al.}(2020)\citenamefont {Sun},
  \citenamefont {Yuan}, \citenamefont {Tsunoda}, \citenamefont {Vedral},
  \citenamefont {Bejamin},\ and\ \citenamefont {Endo}}]{sun2020practical}%
  \BibitemOpen
  \bibfield  {author} {\bibinfo {author} {\bibfnamefont {J.}~\bibnamefont
  {Sun}}, \bibinfo {author} {\bibfnamefont {X.}~\bibnamefont {Yuan}}, \bibinfo
  {author} {\bibfnamefont {T.}~\bibnamefont {Tsunoda}}, \bibinfo {author}
  {\bibfnamefont {V.}~\bibnamefont {Vedral}}, \bibinfo {author} {\bibfnamefont
  {S.~C.}\ \bibnamefont {Bejamin}}, \ and\ \bibinfo {author} {\bibfnamefont
  {S.}~\bibnamefont {Endo}},\ }\href@noop {} {\bibfield  {journal} {\bibinfo
  {journal} {arXiv preprint arXiv:2001.04891}\ } (\bibinfo {year}
  {2020})}\BibitemShut {NoStop}%
\bibitem [{\citenamefont {Endo}\ \emph {et~al.}(2021)\citenamefont {Endo},
  \citenamefont {Cai}, \citenamefont {Benjamin},\ and\ \citenamefont
  {Yuan}}]{endo2021hybrid}%
  \BibitemOpen
  \bibfield  {author} {\bibinfo {author} {\bibfnamefont {S.}~\bibnamefont
  {Endo}}, \bibinfo {author} {\bibfnamefont {Z.}~\bibnamefont {Cai}}, \bibinfo
  {author} {\bibfnamefont {S.~C.}\ \bibnamefont {Benjamin}}, \ and\ \bibinfo
  {author} {\bibfnamefont {X.}~\bibnamefont {Yuan}},\ }\href@noop {} {\bibfield
   {journal} {\bibinfo  {journal} {Journal of the Physical Society of Japan}\
  }\textbf {\bibinfo {volume} {90}},\ \bibinfo {pages} {032001} (\bibinfo
  {year} {2021})}\BibitemShut {NoStop}%
\bibitem [{\citenamefont {Yoshihara}\ \emph {et~al.}(2006)\citenamefont
  {Yoshihara}, \citenamefont {Harrabi}, \citenamefont {Niskanen}, \citenamefont
  {Nakamura},\ and\ \citenamefont {Tsai}}]{yoshihara2006decoherence}%
  \BibitemOpen
  \bibfield  {author} {\bibinfo {author} {\bibfnamefont {F.}~\bibnamefont
  {Yoshihara}}, \bibinfo {author} {\bibfnamefont {K.}~\bibnamefont {Harrabi}},
  \bibinfo {author} {\bibfnamefont {A.~O.}\ \bibnamefont {Niskanen}}, \bibinfo
  {author} {\bibfnamefont {Y.}~\bibnamefont {Nakamura}}, \ and\ \bibinfo
  {author} {\bibfnamefont {J.~S.}\ \bibnamefont {Tsai}},\ }\href@noop {}
  {\bibfield  {journal} {\bibinfo  {journal} {Physical review letters}\
  }\textbf {\bibinfo {volume} {97}},\ \bibinfo {pages} {167001} (\bibinfo
  {year} {2006})}\BibitemShut {NoStop}%
\bibitem [{\citenamefont {Kakuyanagi}\ \emph {et~al.}(2007)\citenamefont
  {Kakuyanagi}, \citenamefont {Meno}, \citenamefont {Saito}, \citenamefont
  {Nakano}, \citenamefont {Semba}, \citenamefont {Takayanagi}, \citenamefont
  {Deppe},\ and\ \citenamefont {Shnirman}}]{kakuyanagi2007dephasing}%
  \BibitemOpen
  \bibfield  {author} {\bibinfo {author} {\bibfnamefont {K.}~\bibnamefont
  {Kakuyanagi}}, \bibinfo {author} {\bibfnamefont {T.}~\bibnamefont {Meno}},
  \bibinfo {author} {\bibfnamefont {S.}~\bibnamefont {Saito}}, \bibinfo
  {author} {\bibfnamefont {H.}~\bibnamefont {Nakano}}, \bibinfo {author}
  {\bibfnamefont {K.}~\bibnamefont {Semba}}, \bibinfo {author} {\bibfnamefont
  {H.}~\bibnamefont {Takayanagi}}, \bibinfo {author} {\bibfnamefont
  {F.}~\bibnamefont {Deppe}}, \ and\ \bibinfo {author} {\bibfnamefont
  {A.}~\bibnamefont {Shnirman}},\ }\href@noop {} {\bibfield  {journal}
  {\bibinfo  {journal} {Physical review letters}\ }\textbf {\bibinfo {volume}
  {98}},\ \bibinfo {pages} {047004} (\bibinfo {year} {2007})}\BibitemShut
  {NoStop}%
\bibitem [{\citenamefont {Bar-Gill}\ \emph {et~al.}(2013)\citenamefont
  {Bar-Gill}, \citenamefont {Pham}, \citenamefont {Jarmola}, \citenamefont
  {Budker},\ and\ \citenamefont {Walsworth}}]{bar2013solid}%
  \BibitemOpen
  \bibfield  {author} {\bibinfo {author} {\bibfnamefont {N.}~\bibnamefont
  {Bar-Gill}}, \bibinfo {author} {\bibfnamefont {L.~M.}\ \bibnamefont {Pham}},
  \bibinfo {author} {\bibfnamefont {A.}~\bibnamefont {Jarmola}}, \bibinfo
  {author} {\bibfnamefont {D.}~\bibnamefont {Budker}}, \ and\ \bibinfo {author}
  {\bibfnamefont {R.~L.}\ \bibnamefont {Walsworth}},\ }\href@noop {} {\bibfield
   {journal} {\bibinfo  {journal} {Nature communications}\ }\textbf {\bibinfo
  {volume} {4}},\ \bibinfo {pages} {1} (\bibinfo {year} {2013})}\BibitemShut
  {NoStop}%
\bibitem [{\citenamefont {De~Lange}\ \emph {et~al.}(2010)\citenamefont
  {De~Lange}, \citenamefont {Wang}, \citenamefont {Riste}, \citenamefont
  {Dobrovitski},\ and\ \citenamefont {Hanson}}]{de2010universal}%
  \BibitemOpen
  \bibfield  {author} {\bibinfo {author} {\bibfnamefont {G.}~\bibnamefont
  {De~Lange}}, \bibinfo {author} {\bibfnamefont {Z.}~\bibnamefont {Wang}},
  \bibinfo {author} {\bibfnamefont {D.}~\bibnamefont {Riste}}, \bibinfo
  {author} {\bibfnamefont {V.}~\bibnamefont {Dobrovitski}}, \ and\ \bibinfo
  {author} {\bibfnamefont {R.}~\bibnamefont {Hanson}},\ }\href@noop {}
  {\bibfield  {journal} {\bibinfo  {journal} {Science}\ }\textbf {\bibinfo
  {volume} {330}},\ \bibinfo {pages} {60} (\bibinfo {year} {2010})}\BibitemShut
  {NoStop}%
\bibitem [{\citenamefont {Kawakami}\ \emph {et~al.}(2014)\citenamefont
  {Kawakami}, \citenamefont {Scarlino}, \citenamefont {Ward}, \citenamefont
  {Braakman}, \citenamefont {Savage}, \citenamefont {Lagally}, \citenamefont
  {Friesen}, \citenamefont {Coppersmith}, \citenamefont {Eriksson},\ and\
  \citenamefont {Vandersypen}}]{kawakami2014electrical}%
  \BibitemOpen
  \bibfield  {author} {\bibinfo {author} {\bibfnamefont {E.}~\bibnamefont
  {Kawakami}}, \bibinfo {author} {\bibfnamefont {P.}~\bibnamefont {Scarlino}},
  \bibinfo {author} {\bibfnamefont {D.~R.}\ \bibnamefont {Ward}}, \bibinfo
  {author} {\bibfnamefont {F.}~\bibnamefont {Braakman}}, \bibinfo {author}
  {\bibfnamefont {D.}~\bibnamefont {Savage}}, \bibinfo {author} {\bibfnamefont
  {M.}~\bibnamefont {Lagally}}, \bibinfo {author} {\bibfnamefont
  {M.}~\bibnamefont {Friesen}}, \bibinfo {author} {\bibfnamefont {S.~N.}\
  \bibnamefont {Coppersmith}}, \bibinfo {author} {\bibfnamefont {M.~A.}\
  \bibnamefont {Eriksson}}, \ and\ \bibinfo {author} {\bibfnamefont
  {L.}~\bibnamefont {Vandersypen}},\ }\href@noop {} {\bibfield  {journal}
  {\bibinfo  {journal} {Nature nanotechnology}\ }\textbf {\bibinfo {volume}
  {9}},\ \bibinfo {pages} {666} (\bibinfo {year} {2014})}\BibitemShut {NoStop}%
\bibitem [{\citenamefont {Watson}\ \emph {et~al.}(2018)\citenamefont {Watson},
  \citenamefont {Philips}, \citenamefont {Kawakami}, \citenamefont {Ward},
  \citenamefont {Scarlino}, \citenamefont {Veldhorst}, \citenamefont {Savage},
  \citenamefont {Lagally}, \citenamefont {Friesen}, \citenamefont {Coppersmith}
  \emph {et~al.}}]{watson2018programmable}%
  \BibitemOpen
  \bibfield  {author} {\bibinfo {author} {\bibfnamefont {T.}~\bibnamefont
  {Watson}}, \bibinfo {author} {\bibfnamefont {S.}~\bibnamefont {Philips}},
  \bibinfo {author} {\bibfnamefont {E.}~\bibnamefont {Kawakami}}, \bibinfo
  {author} {\bibfnamefont {D.}~\bibnamefont {Ward}}, \bibinfo {author}
  {\bibfnamefont {P.}~\bibnamefont {Scarlino}}, \bibinfo {author}
  {\bibfnamefont {M.}~\bibnamefont {Veldhorst}}, \bibinfo {author}
  {\bibfnamefont {D.}~\bibnamefont {Savage}}, \bibinfo {author} {\bibfnamefont
  {M.}~\bibnamefont {Lagally}}, \bibinfo {author} {\bibfnamefont
  {M.}~\bibnamefont {Friesen}}, \bibinfo {author} {\bibfnamefont
  {S.}~\bibnamefont {Coppersmith}},  \emph {et~al.},\ }\href@noop {} {\bibfield
   {journal} {\bibinfo  {journal} {Nature}\ }\textbf {\bibinfo {volume}
  {555}},\ \bibinfo {pages} {633} (\bibinfo {year} {2018})}\BibitemShut
  {NoStop}%
\bibitem [{\citenamefont {Rivas}\ \emph {et~al.}(2014)\citenamefont {Rivas},
  \citenamefont {Huelga},\ and\ \citenamefont {Plenio}}]{rivas2014quantum}%
  \BibitemOpen
  \bibfield  {author} {\bibinfo {author} {\bibfnamefont {A.}~\bibnamefont
  {Rivas}}, \bibinfo {author} {\bibfnamefont {S.~F.}\ \bibnamefont {Huelga}}, \
  and\ \bibinfo {author} {\bibfnamefont {M.~B.}\ \bibnamefont {Plenio}},\
  }\href@noop {} {\bibfield  {journal} {\bibinfo  {journal} {Reports on
  Progress in Physics}\ }\textbf {\bibinfo {volume} {77}},\ \bibinfo {pages}
  {094001} (\bibinfo {year} {2014})}\BibitemShut {NoStop}%
\bibitem [{\citenamefont {Breuer}\ \emph {et~al.}(2016)\citenamefont {Breuer},
  \citenamefont {Laine}, \citenamefont {Piilo},\ and\ \citenamefont
  {Vacchini}}]{breuer2016colloquium}%
  \BibitemOpen
  \bibfield  {author} {\bibinfo {author} {\bibfnamefont {H.-P.}\ \bibnamefont
  {Breuer}}, \bibinfo {author} {\bibfnamefont {E.-M.}\ \bibnamefont {Laine}},
  \bibinfo {author} {\bibfnamefont {J.}~\bibnamefont {Piilo}}, \ and\ \bibinfo
  {author} {\bibfnamefont {B.}~\bibnamefont {Vacchini}},\ }\href@noop {}
  {\bibfield  {journal} {\bibinfo  {journal} {Reviews of Modern Physics}\
  }\textbf {\bibinfo {volume} {88}},\ \bibinfo {pages} {021002} (\bibinfo
  {year} {2016})}\BibitemShut {NoStop}%
\bibitem [{\citenamefont {Breuer}\ \emph {et~al.}(2002)\citenamefont {Breuer},
  \citenamefont {Petruccione} \emph {et~al.}}]{breuer2002theory}%
  \BibitemOpen
  \bibfield  {author} {\bibinfo {author} {\bibfnamefont {H.-P.}\ \bibnamefont
  {Breuer}}, \bibinfo {author} {\bibfnamefont {F.}~\bibnamefont {Petruccione}},
   \emph {et~al.},\ }\href@noop {} {\emph {\bibinfo {title} {The theory of open
  quantum systems}}}\ (\bibinfo  {publisher} {Oxford University Press on
  Demand},\ \bibinfo {year} {2002})\BibitemShut {NoStop}%
\bibitem [{\citenamefont {Wolf}\ and\ \citenamefont
  {Cirac}(2008)}]{wolf2008dividing}%
  \BibitemOpen
  \bibfield  {author} {\bibinfo {author} {\bibfnamefont {M.~M.}\ \bibnamefont
  {Wolf}}\ and\ \bibinfo {author} {\bibfnamefont {J.~I.}\ \bibnamefont
  {Cirac}},\ }\href@noop {} {\bibfield  {journal} {\bibinfo  {journal}
  {Communications in Mathematical Physics}\ }\textbf {\bibinfo {volume}
  {279}},\ \bibinfo {pages} {147} (\bibinfo {year} {2008})}\BibitemShut
  {NoStop}%
\bibitem [{\citenamefont {Rivas}\ \emph {et~al.}(2010)\citenamefont {Rivas},
  \citenamefont {Huelga},\ and\ \citenamefont
  {Plenio}}]{rivas2010entanglement}%
  \BibitemOpen
  \bibfield  {author} {\bibinfo {author} {\bibfnamefont {{\'A}.}~\bibnamefont
  {Rivas}}, \bibinfo {author} {\bibfnamefont {S.~F.}\ \bibnamefont {Huelga}}, \
  and\ \bibinfo {author} {\bibfnamefont {M.~B.}\ \bibnamefont {Plenio}},\
  }\href@noop {} {\bibfield  {journal} {\bibinfo  {journal} {Physical review
  letters}\ }\textbf {\bibinfo {volume} {105}},\ \bibinfo {pages} {050403}
  (\bibinfo {year} {2010})}\BibitemShut {NoStop}%
\bibitem [{\citenamefont {Breuer}\ \emph {et~al.}(2009)\citenamefont {Breuer},
  \citenamefont {Laine},\ and\ \citenamefont {Piilo}}]{breuer2009measure}%
  \BibitemOpen
  \bibfield  {author} {\bibinfo {author} {\bibfnamefont {H.-P.}\ \bibnamefont
  {Breuer}}, \bibinfo {author} {\bibfnamefont {E.-M.}\ \bibnamefont {Laine}}, \
  and\ \bibinfo {author} {\bibfnamefont {J.}~\bibnamefont {Piilo}},\
  }\href@noop {} {\bibfield  {journal} {\bibinfo  {journal} {Physical review
  letters}\ }\textbf {\bibinfo {volume} {103}},\ \bibinfo {pages} {210401}
  (\bibinfo {year} {2009})}\BibitemShut {NoStop}%
\bibitem [{\citenamefont {Misra}\ and\ \citenamefont
  {Sudarshan}(1977)}]{misra1977zeno}%
  \BibitemOpen
  \bibfield  {author} {\bibinfo {author} {\bibfnamefont {B.}~\bibnamefont
  {Misra}}\ and\ \bibinfo {author} {\bibfnamefont {E.~G.}\ \bibnamefont
  {Sudarshan}},\ }\href@noop {} {\bibfield  {journal} {\bibinfo  {journal}
  {Journal of Mathematical Physics}\ }\textbf {\bibinfo {volume} {18}},\
  \bibinfo {pages} {756} (\bibinfo {year} {1977})}\BibitemShut {NoStop}%
\bibitem [{\citenamefont {Itano}\ \emph {et~al.}(1990)\citenamefont {Itano},
  \citenamefont {Heinzen}, \citenamefont {Bollinger},\ and\ \citenamefont
  {Wineland}}]{itano1990quantum}%
  \BibitemOpen
  \bibfield  {author} {\bibinfo {author} {\bibfnamefont {W.~M.}\ \bibnamefont
  {Itano}}, \bibinfo {author} {\bibfnamefont {D.~J.}\ \bibnamefont {Heinzen}},
  \bibinfo {author} {\bibfnamefont {J.~J.}\ \bibnamefont {Bollinger}}, \ and\
  \bibinfo {author} {\bibfnamefont {D.~J.}\ \bibnamefont {Wineland}},\
  }\href@noop {} {\bibfield  {journal} {\bibinfo  {journal} {Physical Review
  A}\ }\textbf {\bibinfo {volume} {41}},\ \bibinfo {pages} {2295} (\bibinfo
  {year} {1990})}\BibitemShut {NoStop}%
\bibitem [{\citenamefont {Viola}\ and\ \citenamefont
  {Lloyd}(1998)}]{viola1998dynamical}%
  \BibitemOpen
  \bibfield  {author} {\bibinfo {author} {\bibfnamefont {L.}~\bibnamefont
  {Viola}}\ and\ \bibinfo {author} {\bibfnamefont {S.}~\bibnamefont {Lloyd}},\
  }\href@noop {} {\bibfield  {journal} {\bibinfo  {journal} {Physical Review
  A}\ }\textbf {\bibinfo {volume} {58}},\ \bibinfo {pages} {2733} (\bibinfo
  {year} {1998})}\BibitemShut {NoStop}%
\bibitem [{\citenamefont {Haikka}\ \emph {et~al.}(2012)\citenamefont {Haikka},
  \citenamefont {Goold}, \citenamefont {McEndoo}, \citenamefont {Plastina},\
  and\ \citenamefont {Maniscalco}}]{haikka2012non}%
  \BibitemOpen
  \bibfield  {author} {\bibinfo {author} {\bibfnamefont {P.}~\bibnamefont
  {Haikka}}, \bibinfo {author} {\bibfnamefont {J.}~\bibnamefont {Goold}},
  \bibinfo {author} {\bibfnamefont {S.}~\bibnamefont {McEndoo}}, \bibinfo
  {author} {\bibfnamefont {F.}~\bibnamefont {Plastina}}, \ and\ \bibinfo
  {author} {\bibfnamefont {S.}~\bibnamefont {Maniscalco}},\ }\href@noop {}
  {\bibfield  {journal} {\bibinfo  {journal} {Physical Review A}\ }\textbf
  {\bibinfo {volume} {85}},\ \bibinfo {pages} {060101} (\bibinfo {year}
  {2012})}\BibitemShut {NoStop}%
\bibitem [{\citenamefont {Vasile}\ \emph {et~al.}(2011)\citenamefont {Vasile},
  \citenamefont {Olivares}, \citenamefont {Paris},\ and\ \citenamefont
  {Maniscalco}}]{vasile2011continuous}%
  \BibitemOpen
  \bibfield  {author} {\bibinfo {author} {\bibfnamefont {R.}~\bibnamefont
  {Vasile}}, \bibinfo {author} {\bibfnamefont {S.}~\bibnamefont {Olivares}},
  \bibinfo {author} {\bibfnamefont {M.~G.~A.}\ \bibnamefont {Paris}}, \ and\
  \bibinfo {author} {\bibfnamefont {S.}~\bibnamefont {Maniscalco}},\
  }\href@noop {} {\bibfield  {journal} {\bibinfo  {journal} {Physical Review
  A}\ }\textbf {\bibinfo {volume} {83}},\ \bibinfo {pages} {042321} (\bibinfo
  {year} {2011})}\BibitemShut {NoStop}%
\bibitem [{\citenamefont {Haikka}\ \emph {et~al.}(2013)\citenamefont {Haikka},
  \citenamefont {Johnson},\ and\ \citenamefont {Maniscalco}}]{haikka2013non}%
  \BibitemOpen
  \bibfield  {author} {\bibinfo {author} {\bibfnamefont {P.}~\bibnamefont
  {Haikka}}, \bibinfo {author} {\bibfnamefont {T.}~\bibnamefont {Johnson}}, \
  and\ \bibinfo {author} {\bibfnamefont {S.}~\bibnamefont {Maniscalco}},\
  }\href@noop {} {\bibfield  {journal} {\bibinfo  {journal} {Physical Review
  A}\ }\textbf {\bibinfo {volume} {87}},\ \bibinfo {pages} {010103} (\bibinfo
  {year} {2013})}\BibitemShut {NoStop}%
\bibitem [{\citenamefont {{\v{Z}}nidari{\v{c}}}\ \emph
  {et~al.}(2011)\citenamefont {{\v{Z}}nidari{\v{c}}}, \citenamefont {Pineda},\
  and\ \citenamefont {Garcia-Mata}}]{vznidarivc2011non}%
  \BibitemOpen
  \bibfield  {author} {\bibinfo {author} {\bibfnamefont {M.}~\bibnamefont
  {{\v{Z}}nidari{\v{c}}}}, \bibinfo {author} {\bibfnamefont {C.}~\bibnamefont
  {Pineda}}, \ and\ \bibinfo {author} {\bibfnamefont {I.}~\bibnamefont
  {Garcia-Mata}},\ }\href@noop {} {\bibfield  {journal} {\bibinfo  {journal}
  {Physical review letters}\ }\textbf {\bibinfo {volume} {107}},\ \bibinfo
  {pages} {080404} (\bibinfo {year} {2011})}\BibitemShut {NoStop}%
\bibitem [{\citenamefont {Wakakuwa}(2017)}]{wakakuwa2017operational}%
  \BibitemOpen
  \bibfield  {author} {\bibinfo {author} {\bibfnamefont {E.}~\bibnamefont
  {Wakakuwa}},\ }\href@noop {} {\bibfield  {journal} {\bibinfo  {journal}
  {arXiv preprint arXiv:1709.07248}\ } (\bibinfo {year} {2017})}\BibitemShut
  {NoStop}%
\bibitem [{\citenamefont {Matsuzaki}\ \emph {et~al.}(2011)\citenamefont
  {Matsuzaki}, \citenamefont {Benjamin},\ and\ \citenamefont
  {Fitzsimons}}]{matsuzaki2011magnetic}%
  \BibitemOpen
  \bibfield  {author} {\bibinfo {author} {\bibfnamefont {Y.}~\bibnamefont
  {Matsuzaki}}, \bibinfo {author} {\bibfnamefont {S.~C.}\ \bibnamefont
  {Benjamin}}, \ and\ \bibinfo {author} {\bibfnamefont {J.}~\bibnamefont
  {Fitzsimons}},\ }\href@noop {} {\bibfield  {journal} {\bibinfo  {journal}
  {Physical Review A}\ }\textbf {\bibinfo {volume} {84}},\ \bibinfo {pages}
  {012103} (\bibinfo {year} {2011})}\BibitemShut {NoStop}%
\bibitem [{\citenamefont {Chin}\ \emph {et~al.}(2012)\citenamefont {Chin},
  \citenamefont {Huelga},\ and\ \citenamefont {Plenio}}]{chin2012quantum}%
  \BibitemOpen
  \bibfield  {author} {\bibinfo {author} {\bibfnamefont {A.~W.}\ \bibnamefont
  {Chin}}, \bibinfo {author} {\bibfnamefont {S.~F.}\ \bibnamefont {Huelga}}, \
  and\ \bibinfo {author} {\bibfnamefont {M.~B.}\ \bibnamefont {Plenio}},\
  }\href@noop {} {\bibfield  {journal} {\bibinfo  {journal} {Physical review
  letters}\ }\textbf {\bibinfo {volume} {109}},\ \bibinfo {pages} {233601}
  (\bibinfo {year} {2012})}\BibitemShut {NoStop}%
\bibitem [{\citenamefont {Hall}\ \emph {et~al.}(2014)\citenamefont {Hall},
  \citenamefont {Cresser}, \citenamefont {Li},\ and\ \citenamefont
  {Andersson}}]{hall2014canonical}%
  \BibitemOpen
  \bibfield  {author} {\bibinfo {author} {\bibfnamefont {M.~J.~W.}\
  \bibnamefont {Hall}}, \bibinfo {author} {\bibfnamefont {J.~D.}\ \bibnamefont
  {Cresser}}, \bibinfo {author} {\bibfnamefont {L.}~\bibnamefont {Li}}, \ and\
  \bibinfo {author} {\bibfnamefont {E.}~\bibnamefont {Andersson}},\ }\href@noop
  {} {\bibfield  {journal} {\bibinfo  {journal} {Physical Review A}\ }\textbf
  {\bibinfo {volume} {89}},\ \bibinfo {pages} {042120} (\bibinfo {year}
  {2014})}\BibitemShut {NoStop}%
\bibitem [{\citenamefont {Binho}\ \emph {et~al.}(2019)\citenamefont {Binho},
  \citenamefont {Matsuzaki}, \citenamefont {Matsuzaki}, \citenamefont {Kondo}
  \emph {et~al.}}]{matsuzaki2019realization}%
  \BibitemOpen
  \bibfield  {author} {\bibinfo {author} {\bibfnamefont {L.}~\bibnamefont
  {Binho}}, \bibinfo {author} {\bibfnamefont {Y.}~\bibnamefont {Matsuzaki}},
  \bibinfo {author} {\bibfnamefont {M.}~\bibnamefont {Matsuzaki}}, \bibinfo
  {author} {\bibfnamefont {Y.}~\bibnamefont {Kondo}},  \emph {et~al.},\
  }\href@noop {} {\bibfield  {journal} {\bibinfo  {journal} {New Journal of
  Physics}\ }\textbf {\bibinfo {volume} {21}},\ \bibinfo {pages} {093008}
  (\bibinfo {year} {2019})}\BibitemShut {NoStop}%
\bibitem [{\citenamefont {Kukita}\ \emph {et~al.}(2020)\citenamefont {Kukita},
  \citenamefont {Kondo},\ and\ \citenamefont {Nakahara}}]{Kukita_2020}%
  \BibitemOpen
  \bibfield  {author} {\bibinfo {author} {\bibfnamefont {S.}~\bibnamefont
  {Kukita}}, \bibinfo {author} {\bibfnamefont {Y.}~\bibnamefont {Kondo}}, \
  and\ \bibinfo {author} {\bibfnamefont {M.}~\bibnamefont {Nakahara}},\ }\href
  {\doibase 10.1088/1367-2630/abbfcf} {\bibfield  {journal} {\bibinfo
  {journal} {New Journal of Physics}\ }\textbf {\bibinfo {volume} {22}},\
  \bibinfo {pages} {103048} (\bibinfo {year} {2020})}\BibitemShut {NoStop}%
\bibitem [{\citenamefont {Govenius}\ \emph {et~al.}(2015)\citenamefont
  {Govenius}, \citenamefont {Matsuzaki}, \citenamefont {Savenko},\ and\
  \citenamefont {M{\"o}tt{\"o}nen}}]{govenius2015parity}%
  \BibitemOpen
  \bibfield  {author} {\bibinfo {author} {\bibfnamefont {J.}~\bibnamefont
  {Govenius}}, \bibinfo {author} {\bibfnamefont {Y.}~\bibnamefont {Matsuzaki}},
  \bibinfo {author} {\bibfnamefont {I.~G.}\ \bibnamefont {Savenko}}, \ and\
  \bibinfo {author} {\bibfnamefont {M.}~\bibnamefont {M{\"o}tt{\"o}nen}},\
  }\href@noop {} {\bibfield  {journal} {\bibinfo  {journal} {Physical Review
  A}\ }\textbf {\bibinfo {volume} {92}},\ \bibinfo {pages} {042305} (\bibinfo
  {year} {2015})}\BibitemShut {NoStop}%
\bibitem [{\citenamefont {Blais}\ \emph {et~al.}(2004)\citenamefont {Blais},
  \citenamefont {Huang}, \citenamefont {Wallraff}, \citenamefont {Girvin},\
  and\ \citenamefont {Schoelkopf}}]{blais2004cavity}%
  \BibitemOpen
  \bibfield  {author} {\bibinfo {author} {\bibfnamefont {A.}~\bibnamefont
  {Blais}}, \bibinfo {author} {\bibfnamefont {R.-S.}\ \bibnamefont {Huang}},
  \bibinfo {author} {\bibfnamefont {A.}~\bibnamefont {Wallraff}}, \bibinfo
  {author} {\bibfnamefont {S.~M.}\ \bibnamefont {Girvin}}, \ and\ \bibinfo
  {author} {\bibfnamefont {R.~J.}\ \bibnamefont {Schoelkopf}},\ }\href@noop {}
  {\bibfield  {journal} {\bibinfo  {journal} {Physical Review A}\ }\textbf
  {\bibinfo {volume} {69}},\ \bibinfo {pages} {062320} (\bibinfo {year}
  {2004})}\BibitemShut {NoStop}%
\bibitem [{\citenamefont {Bertet}\ \emph {et~al.}(2005)\citenamefont {Bertet},
  \citenamefont {Chiorescu}, \citenamefont {Burkard}, \citenamefont {Semba},
  \citenamefont {Harmans}, \citenamefont {DiVincenzo},\ and\ \citenamefont
  {Mooij}}]{bertet2005dephasing}%
  \BibitemOpen
  \bibfield  {author} {\bibinfo {author} {\bibfnamefont {P.}~\bibnamefont
  {Bertet}}, \bibinfo {author} {\bibfnamefont {I.}~\bibnamefont {Chiorescu}},
  \bibinfo {author} {\bibfnamefont {G.}~\bibnamefont {Burkard}}, \bibinfo
  {author} {\bibfnamefont {K.}~\bibnamefont {Semba}}, \bibinfo {author}
  {\bibfnamefont {C.~J. P.~M.}\ \bibnamefont {Harmans}}, \bibinfo {author}
  {\bibfnamefont {D.~P.}\ \bibnamefont {DiVincenzo}}, \ and\ \bibinfo {author}
  {\bibfnamefont {J.~E.}\ \bibnamefont {Mooij}},\ }\href@noop {} {\bibfield
  {journal} {\bibinfo  {journal} {Physical review letters}\ }\textbf {\bibinfo
  {volume} {95}},\ \bibinfo {pages} {257002} (\bibinfo {year}
  {2005})}\BibitemShut {NoStop}%
\bibitem [{\citenamefont {de~Vega}\ and\ \citenamefont
  {Alonso}(2017)}]{de2017dynamics}%
  \BibitemOpen
  \bibfield  {author} {\bibinfo {author} {\bibfnamefont {I.}~\bibnamefont
  {de~Vega}}\ and\ \bibinfo {author} {\bibfnamefont {D.}~\bibnamefont
  {Alonso}},\ }\href@noop {} {\bibfield  {journal} {\bibinfo  {journal}
  {Reviews of Modern Physics}\ }\textbf {\bibinfo {volume} {89}},\ \bibinfo
  {pages} {015001} (\bibinfo {year} {2017})}\BibitemShut {NoStop}%
\bibitem [{\citenamefont {Choi}(1975)}]{choi1975completely}%
  \BibitemOpen
  \bibfield  {author} {\bibinfo {author} {\bibfnamefont {M.-D.}\ \bibnamefont
  {Choi}},\ }\href@noop {} {\bibfield  {journal} {\bibinfo  {journal} {Linear
  algebra and its applications}\ }\textbf {\bibinfo {volume} {10}},\ \bibinfo
  {pages} {285} (\bibinfo {year} {1975})}\BibitemShut {NoStop}%
\bibitem [{\citenamefont {Jamio{\l}kowski}(1972)}]{jamiolkowski1972linear}%
  \BibitemOpen
  \bibfield  {author} {\bibinfo {author} {\bibfnamefont {A.}~\bibnamefont
  {Jamio{\l}kowski}},\ }\href@noop {} {\bibfield  {journal} {\bibinfo
  {journal} {Reports on Mathematical Physics}\ }\textbf {\bibinfo {volume}
  {3}},\ \bibinfo {pages} {275} (\bibinfo {year} {1972})}\BibitemShut {NoStop}%
\bibitem [{\citenamefont {Greenbaum}(2015)}]{greenbaum2015introduction}%
  \BibitemOpen
  \bibfield  {author} {\bibinfo {author} {\bibfnamefont {D.}~\bibnamefont
  {Greenbaum}},\ }\href@noop {} {\bibfield  {journal} {\bibinfo  {journal}
  {arXiv preprint arXiv:1509.02921}\ } (\bibinfo {year} {2015})}\BibitemShut
  {NoStop}%
\bibitem [{\citenamefont {Urrego}\ \emph {et~al.}(2018)\citenamefont {Urrego},
  \citenamefont {Fl\'orez}, \citenamefont {Svozil\'{\i}k}, \citenamefont
  {Nu\~nez},\ and\ \citenamefont {Valencia}}]{PhysRevA.98.053862}%
  \BibitemOpen
  \bibfield  {author} {\bibinfo {author} {\bibfnamefont {D.~F.}\ \bibnamefont
  {Urrego}}, \bibinfo {author} {\bibfnamefont {J.}~\bibnamefont {Fl\'orez}},
  \bibinfo {author} {\bibfnamefont {J.~c.~v.}\ \bibnamefont {Svozil\'{\i}k}},
  \bibinfo {author} {\bibfnamefont {M.}~\bibnamefont {Nu\~nez}}, \ and\
  \bibinfo {author} {\bibfnamefont {A.}~\bibnamefont {Valencia}},\ }\href
  {\doibase 10.1103/PhysRevA.98.053862} {\bibfield  {journal} {\bibinfo
  {journal} {Phys. Rev. A}\ }\textbf {\bibinfo {volume} {98}},\ \bibinfo
  {pages} {053862} (\bibinfo {year} {2018})}\BibitemShut {NoStop}%
\bibitem [{\citenamefont {Valente}\ \emph {et~al.}(2016)\citenamefont
  {Valente}, \citenamefont {Arruda},\ and\ \citenamefont
  {Werlang}}]{Valente:16}%
  \BibitemOpen
  \bibfield  {author} {\bibinfo {author} {\bibfnamefont {D.}~\bibnamefont
  {Valente}}, \bibinfo {author} {\bibfnamefont {M.~F.~Z.}\ \bibnamefont
  {Arruda}}, \ and\ \bibinfo {author} {\bibfnamefont {T.}~\bibnamefont
  {Werlang}},\ }\href {\doibase 10.1364/OL.41.003126} {\bibfield  {journal}
  {\bibinfo  {journal} {Opt. Lett.}\ }\textbf {\bibinfo {volume} {41}},\
  \bibinfo {pages} {3126} (\bibinfo {year} {2016})}\BibitemShut {NoStop}%
\end{thebibliography}%

\end{document}